\documentclass[preprint,prb,preprintnumbers,showpacs,amsmath,amssymb,superscriptaddress]{revtex4}

\usepackage{graphicx}
\usepackage{dcolumn}
\usepackage{bm}

\begin{document}

\preprint{}

\title{Observation of the spin-charge thermal isolation of ferromagnetic $\bf{Ga_{0.94}Mn_{0.06}As}$
 by time-resolved magneto-optical measurement}

\author{E.Kojima}
\affiliation{Department of Applied Physics, The University of Tokyo, Tokyo 113-8656, Japan}
\affiliation{Cooperative Excitation Project, ERATO, Japan Science and Technology Corporation(JST), Kanagawa 213-0012, Japan}
\author{R.Shimano}
\affiliation{Department of Applied Physics, The University of Tokyo, Tokyo 113-8656, Japan}
\author{Y.Hashimoto}
\affiliation{Institute for Solid State Physics, University of Tokyo, 6-3-7 Kashiwanoha, Kashiwa, Chiba 277-8581, Japan}
\author{S.Katsumoto}
\affiliation{Institute for Solid State Physics, University of Tokyo, 6-3-7 Kashiwanoha, Kashiwa, Chiba 277-8581, Japan}
\author{Y.Iye}
\affiliation{Institute for Solid State Physics, University of Tokyo, 6-3-7 Kashiwanoha, Kashiwa, Chiba 277-8581, Japan}
\author{M.Kuwata-Gonokami}
\email[To whom correspondence should be addressed. E-mail:]{ gonokami@ap.t.u-tokyo.ac.jp}
\affiliation{Department of Applied Physics, The University of Tokyo, Tokyo 113-8656, Japan}
\affiliation{Cooperative Excitation Project, ERATO, Japan Science and Technology Corporation(JST), Kanagawa 213-0012, Japan}

\date{\today}

\begin{abstract}
The dynamics of magnetization under femtosecond optical excitation is studied 
in a ferromagnetic semiconductor ${\rm Ga_{0.94}Mn_{0.06}As}$ with 
a time-resolved magneto-optical Kerr effect measurement with two color probe beams. 
The transient reflectivity change indicates the rapid rise of the carrier temperature 
and relaxation to a quasi-thermal equilibrium within a few ps, while a very slow rise of 
the spin temperature of the order of 500ps is observed. 
This anomalous behavior originates from the thermal isolation 
between the charge and spin systems due to the spin polarization of 
carriers (holes) contributing to ferromagnetism. 
This constitutes experimental proof of the half-metallic nature of 
ferromagnetic ${\rm Ga_{0.94}Mn_{0.06}As}$ 
arising from the large exchange energy of carriers compared to Fermi energy.
\end{abstract}
\maketitle
Carrier-induced ferromagnetism is one of the most visited spontaneous symmetry-breaking phenomena in the solid-state physics. 
The underlying physics of this phenomenon deeply involves fundamental aspects of quantum physics including interactions between localized and delocalized carriers, where many-body quantum correlation plays an essential role. 

The discovery of diluted magnetic semiconductors (DMS's) with high Curie temperature that have become available due to recent advances in molecular beam epitaxy and transition metal ion doping techniques opens new horizons in exploring of the carrier mediated ferromagnetism\cite{Munekata,HOhno1,HOhno2}. 
In particular, the compatibility of DMS's with conventional semiconductor technology is very promising to employ the spin degree of freedom in practical devices. Correspondingly, along with extensive efforts to clarify the microscopic mechanism of the ferromagnetism, 
much work has been devoted to the search for systems with higher $T_c$\cite{Matsukura,Akai,Dietl,Inoue,HOhno3}.

In the initial stage of the research for the origin of the ferromagnetism, two mechanisms of the interaction between the local moments were considered. The first was based on the Ruderman-Kittel-Kasuya-Yoshida (RKKY) mechanism, in which the ferromagnetic interaction of localized spins is mediated by nearly free carriers \cite{Matsukura}. This model reproduces the observed concentration dependence of Curie temperatures in II-VI DMS\cite{Haury}. 

However, recent photoemission experiments in Mn doped GaAs have revealed a rather low density of state, which indicates localization of carriers\cite{Okabayashi} and thus we need to consider the mechanism beyond simple RKKY model.
The second one was based on the double exchange mechanism which is suggested by a band calculation of In$_{1-x}$Mn$_x$As \cite{Akai}. 
In this approach, the ferromagnetic coupling between localized spins is induced by the presence of carriers with strong d-band like character. These carriers originate from transition
metal impurities and essentially contribute to the conductivity of the material. This model was also confronted with the difficulty from both the photoemission and soft X-ray absorption data\cite{Ishiwata}, which indicate that the holes have very little d-component.
Hence at present the researchers are looking for the third way which compromises the above two approach and many hypotheses have been proposed.

The pronounced spin polarization of conducting carriers 
in several ferromagnetic materials, which are often referred to as half-metals 
in which the density of state is nonzero for only one spin direction at the Fermi surface, makes these materials very attractive for spintronics applications \cite{Wolf}. Among high-$T_c$ ferromagnetic semiconductors, half-metallic nature have been suggested for the Mn doped GaAs. Such a suggestion is supported by results of the polarized electro-luminescence \cite{YOhno} and tunneling magneto-resistance measurements\cite{Tanaka} along with {\it ab initio} calculation of the band structure of this material \cite{Jain}. The half-metallicity is very important not only for the application but also for putting a strong constraint for the theories to explain the ferromagnetism. Nevertheless, since the exchange interaction energy is of the same order as Fermi and potential fluctuation energy, the existing experimental data do not allow 
us to arrive at decisive conclusion of the electronic structure of Mn doped GaAs and their half-metallicity.

Simultaneous probing of the spin and carrier dynamics by time-resolved magneto-optical Kerr effect (TR-MOKE) measurements gives us a chance to reveal the half-metallic nature of the material.
The TR-MOKE technique has been first applied to the investigation of the picosecond photo-induced demagnetization of Ni\cite{Beaurepaire}. 
The phenomenon is analyzed by \textquotedblleft three-temperature model \textquotedblright which describes the dynamical process of thermal interaction among sub-systems(charge, spin, and lattice systems).
We have employed the TR-MOKE technique to investigate the photo-induced spin and charge dynamics in ordered double perovskite Sr$_2$FeMoO$_6$\cite{Kise}, which is known to show very large magneto-resistance at room temperature. In particular, we have discovered that the response of the spin system is much slower than the charge system and shown that such a spin-charge thermal isolation originates from the half-metallic nature of Sr$_2$FeMoO$_6$. In turn, the observation of the spin-charge thermal isolation may be suggested as an experimental criterion of the half-metallic nature of a material. 

Although TR-MOKE is an unique method to probe ultrafast dynamics
of magnetization, the extraction of the pure magnetization
component from the MOKE signal is an important
experimental problem because a photo-induced changes of refractive index also contributes to the pump-induced TR-MOKE signal. In the experiment with ferromagnetic metal films, the magnetization component of the signal has been extracted by comparing the results of the linear and nonlinear magneto-optical measurements in both reflection and transmission geometries \cite{Guidoni}.

In this paper, we propose and develop a two color probe method of TR-MOKE applicable to opaque DMS's, which allows us to extract magnetization component in reflection geometry. With this method, we investigate magnetization dynamics in ${\rm Ga_{0.94}Mn_{0.06}As}$ which has the highest Curie temperature, $T_c$=110K among ferromagnetic semiconductors\cite{HOhno3}. The extracted magnetization dynamics reveals that spins thermalize much slowly in comparison with the holes indicating the half-metallicity of this material, supporting some band calculations \cite{Jain} based on a large exchange energy.
A sample of ${\rm Ga_{0.94}Mn_{0.06}As}$ with 1.05${\rm \mu m}$ thickness was grown by molecular beam epitaxy on a GaAs[001] substrate and 
annealed at 280$^\circ$C.

The measurements of linear spectra and TR-MOKE were performed in the polar Kerr configuration under a magnetic field of 0.2T by alternating the field direction in a liquid N$_2$ flow cryostat. Detailed description of the experiment was shown in the previous paper\cite{Kise}.
In the TR-MOKE measurements, a Ti-sapphire regenerative amplifier system with a photon energy of 1.55eV, a pulse duration of 150 fs, and a repetition rate of 1 KHz was used as a light source. The pulses at the fundamental and doubled frequencies were used as a probe and a pump, respectively. The intensity of the pump pulse was 45${\rm \mu J/cm^2}$, which corresponds to the carrier density of 6.2$\times$10$^{19}{\rm cm}^{-3}$. We also performed a TR-MOKE measurement at different probe photon energy under the same pump condition. Part of the white light continuum obtained with sapphire crystal was spectrally filtered and used for the probe pulse with 1.77eV photon energy. The stability of the light source limits the sensitivity of our measurements, typically 1 mdeg at 1.77eV, and 0.2mdeg at 1.55eV. The temporal resolution of 0.3ps was limited by the pump pulse duration.

\begin{figure}[h!]
\includegraphics{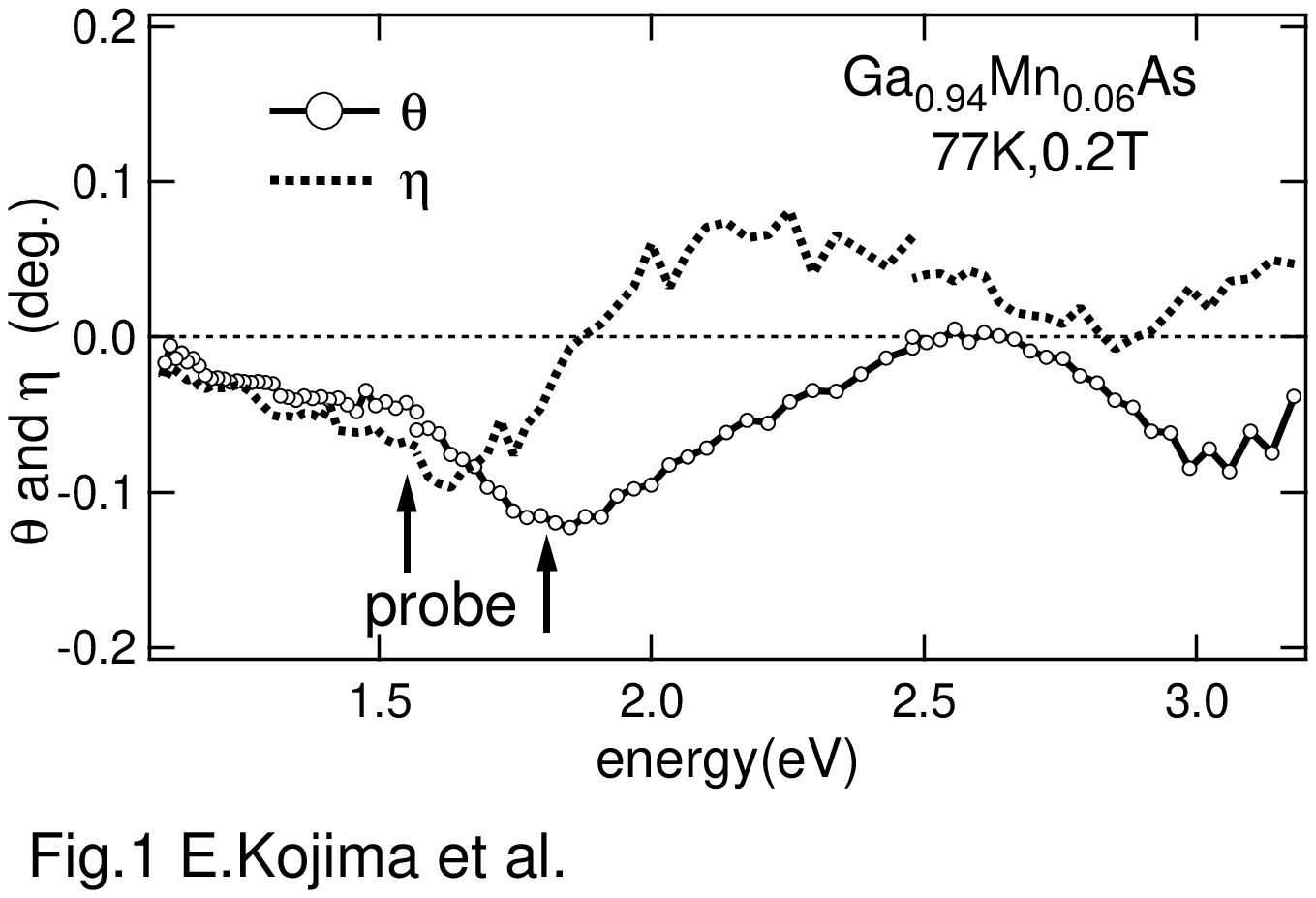}
\caption{\label{fig1} Magneto-optical Kerr spectrum on ${\rm Ga_{0.94}Mn_{0.06}As}$ under a magnetic 
field 0.2T at 77K. The open circles show Kerr rotation. The dotted line shows the Kerr 
ellipticity. Two arrows show the photon energy of the probe pulse used in the TR-MOKE 
measurement.}
\end{figure}

\begin{figure}[h!]
\resizebox{!}{18cm}{\includegraphics{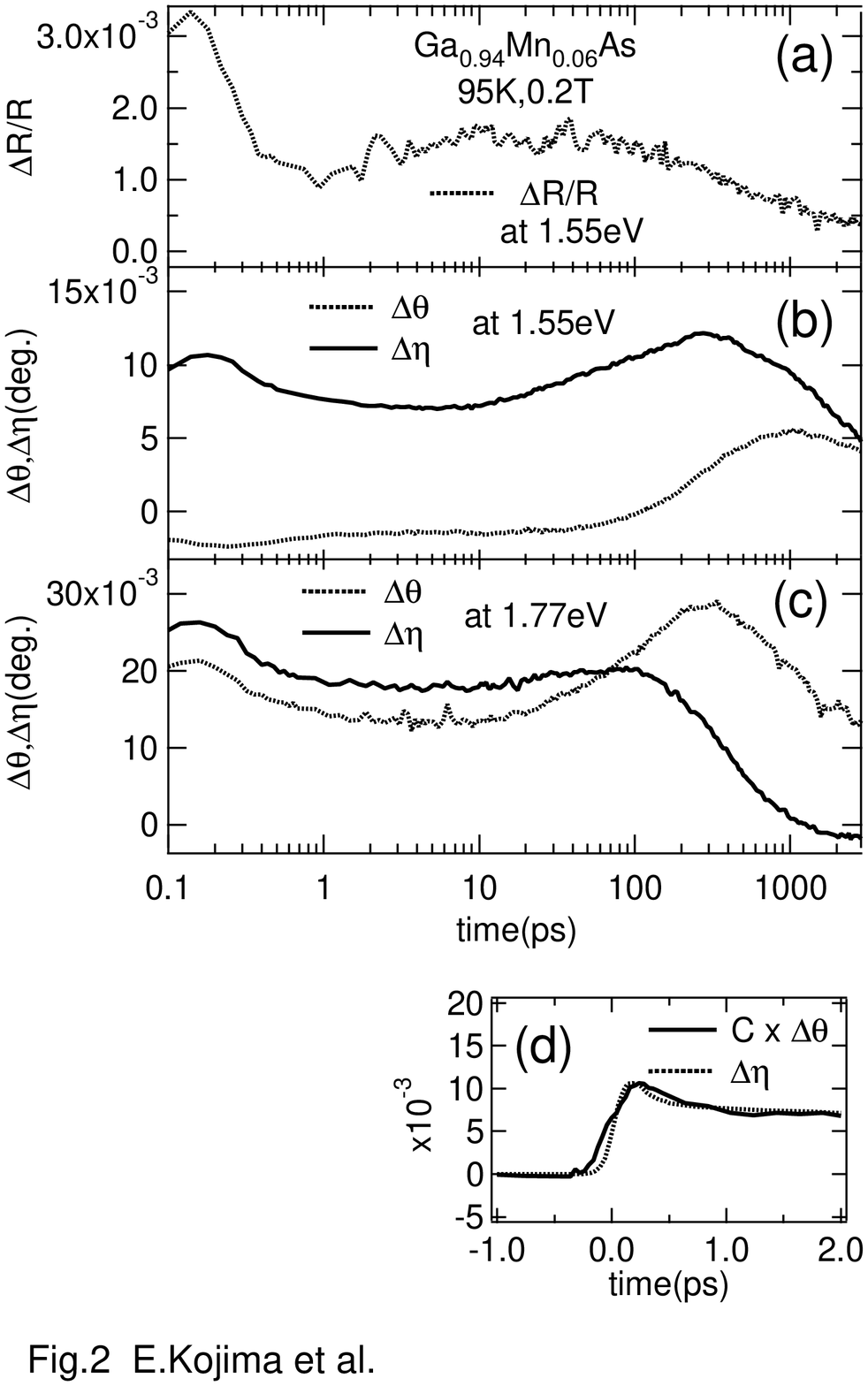}}
\caption{\label{fig2} Temporal evolution of reflectivity change $\Delta R/R$(a) and 
Kerr rotation $\Delta \theta$ and ellipticity $\Delta \eta$ at 1.55eV(b) and 1.77eV(c) 
probe photon energy, 
respectively, under a magnetic field of 0.2 T at 95K. The pump photon energy is 3.1eV. 
The pump fluence is 45${\rm \mu J/cm^2}$. A small inset (d) shows $\Delta \theta$ multiplied by some constant 
and $\Delta \eta$ in the initial process at 1.55eV}
\end{figure}

Figure~\ref{fig1} shows the Kerr rotation and ellipticity spectrum at 77K. 
The arrows in Fig.~\ref{fig1} indicate the photon energies of the probe pulses (1.55eV and 1.77eV) used in the TR-MOKE measurement. Figure~\ref{fig2}(a) shows the dependence of the differential reflectivity at 1.55eV under a 0.2T magnetic field at 95K on the time delay between the pump and probe pulses. 
One can observe a resolution-limited rapid reflectivity change, while a very slow relaxation is observed after several 100 ps.
The temporal evolution of the Kerr rotation $\Delta \theta$ and ellipticity $\Delta \eta$ at 1.55eV and 1.77eV (Figures~\ref{fig2}(b) and (c), respectively) are more complicated in comparison with that of 
Sr$_2$FeMoO$_6$\cite{Kise}. In particular, $\Delta \theta$ at 1.55eV is initially negative, however it changes its sign at later time. 
Furthermore, the strong dependence of the temporal profiles on the probe frequency clearly shows that TR-MOKE signals do not 
directly give information of magnetization. 
In order to analyze these signals, let us present the Kerr rotation and ellipticity in the following form: 
\begin{equation}
\theta=f_\theta \cdot M,\eta=f_\eta \cdot M
\end{equation}
where $M$ is the magnetization, while $f_\theta$ and $f_\eta$ depend on the electronic properties of the 
material and can be presented in terms of the refractive index and absorption coefficient. 
Correspondingly, the photo-induced change in $\theta$ and $\eta$ consists of two components: 
\begin{equation}
\begin{array}{c}	
\Delta \theta(t) \approx f_\theta \cdot \Delta M(t)+\Delta f_\theta (t) \cdot M\\
\Delta \eta(t) \approx f_\eta \cdot \Delta M(t)+\Delta f_\eta (t) \cdot M
\end{array}
\end{equation}
At the probe frequencies of the present experiment, 
$\theta<0$ and $\eta<0$(see Fig.~\ref{fig1}), i.e. 
both coefficients, $f_\theta<0$ and $f_\eta<0$.
Thus, the reduction of 
magnetization leads the positive signal for both $\Delta \theta$ and $\Delta \eta$ from the 
first terms of Eq.(2). 
The second terms, $\Delta f_\theta$ and $\Delta f_\eta$ are the photo-induced change of Kerr signals associated 
with the complex refractive index change($\Delta \tilde{n}$) caused by photo-excitation.
At a probe energy of 1.55 eV (Fig.~\ref{fig2} (b)), we observe $\Delta \theta<0$ and $\Delta \eta>0$ right after the excitation.
This indicates that the rapid changes in $\theta$ and $\eta$ are governed by the second term of Eq.(2).

\begin{figure}[h!]
\includegraphics{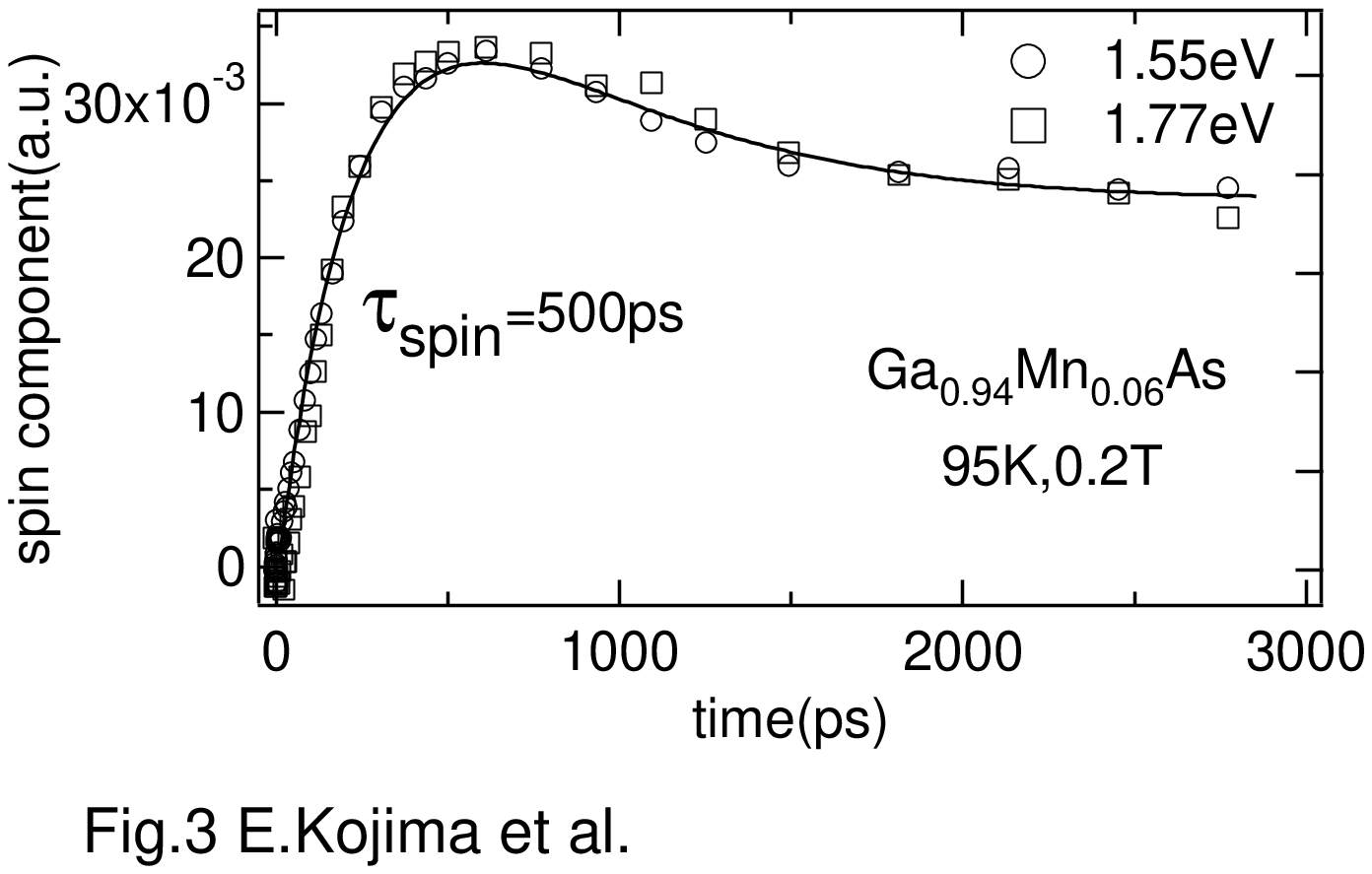}
\caption{\label{fig3}Temporal evolution of the extracted spin component 
(the open circles are from $\Delta \theta$ and $\Delta \eta$ at 1.55eV, and the open 
squares are 
from $\Delta \theta$ and $\Delta \eta$ at 1.77eV). The open circles and squares are 
normalized at 
the maximum point. A positive direction shows demagnetization. 
A solid line shows a result obtained 
using the three-temperature model considering thermal diffusion.}
\end{figure}

Fig.~\ref{fig2}(d) presents temporal profiles of $\Delta \theta$ and $\Delta \eta$ at probe 1.55eV (Fig.~\ref{fig2}(b)) 
within the 2 ps time interval. By multiplying $\Delta \theta$ by -4.6, both curves fit well. 
This indicates that in this time interval, both $\Delta \theta$ and $\Delta \eta$ are governed by a single dynamical parameter.
Therefore, we can take $\Delta f_\eta (t)= C \Delta f_\theta (t)$ where $C$ is a constant in the whole 
temporal region. By substituting this equation into Eqs. (2) we readily arrive at the conclusion that 
the value of $C \Delta \theta-\Delta \eta$ is proportional to the magnetization change $\Delta M(t)$: 
\begin{equation}
(C f_\theta-f_\eta)\Delta M(t)=C\Delta \theta (t)-\Delta \eta(t)
\end{equation}
Eq.(3) shows that we can extract $\Delta M$ from the measured values of MOKE and TR-MOKE signals and  parameter $C$, which depend on the probe frequency\cite{aboutC}. To examine the validity of this procedure, we analyzed the data for the probe at 1.77eV to obtain parameter $C$ and extracted $\Delta M$.
Figure~\ref{fig3} shows, that the normalized $C \Delta \theta-\Delta \eta$ at 1.55eV (open circles) and 
1.77eV (open squares) show the same temporal behavior. The chosen values of $C$ are -4.6 and 1.2 for 1.55eV and 1.77eV, respectively. 
Our results clearly show that we can extract 
the magnetization component from TR-MOKE signals.

Now we discuss the temporal responses of $\Delta R/R$ shown in  Fig.~\ref{fig2}(a) and $\Delta f_\theta$ and $\Delta f_\eta$ of eq.(2). In semiconductors, these changes are driven by the photo-induced change in the complex refractive index caused by the band filling effects and successive carrier heating.
In ${\rm Ga_{1-x}Mn_xAs}$ as well as other low-temperature grown GaAs, the excess arsenic atoms form 
high density As antisites ($\sim 10^{19}$cm$^{-3}$), which act as  deep donor levels. 
The photo-generated electrons are immediately trapped in such levels and non-radiatively recombined 
within 1ps\cite{Benjamin}. It is known that the additional acceptor doping lowers the Fermi level and 
promotes the ionization of deep donor levels, which enhances electron trapping. In the present 
experiment, the high density Mn doping brings acceptor levels with density as high as 10$^{19}$cm$^{-3}$. 
Thus the trapping and non-radiative recombination of photo-excited carriers should be very efficient and the carrier lifetime is less than 1ps\cite{Stellmacher,Haiml}. The resolution-limited fast decay of $\Delta R/R$ within 1ps corresponds to such a rapid non-radiative recombination process. 
The non-radiative process heats up the lattice and the preexisting holes. 
Comparing with the temperature dependence of the linear reflectivity spectrum, we estimate the temperature increase of the holes after the fast decay
as about 3-5 K. 
Considering the small heat capacity of holes, which is four order smaller than that of the lattice, the estimated small temperature increase of holes indicates that holes and lattice 
should have reached a thermal equilibrium during the non-radiative recombination processes. 
The slow decay observed in $\Delta R/R$ over several hundred ps is due to thermal diffusion.

After the non-radiative recombination, $\Delta \tilde{n}$ induced by the temperature increase of holes governs the temporal responses of $\Delta R/R$ and $\Delta f_\theta$ and $\Delta f_\eta$ as we have noted that all the signals are described with a single dynamical parameter shown in Fig.~\ref{fig2}(d). 

Now we discuss the origin of the slow thermalization of the extracted magnetization in Fig.~\ref{fig3}.
The observed slow thermalization of spin systems is rather unusual for GaAs based semiconductors 
because the valence band is strongly affected by spin-orbit interaction and the heavy hole and light hole 
bands are degenerated at the $\Gamma$ point. 
The spin relaxation time for the hole in pure GaAs is known to be about 110fs \cite{Hilton}. 
Moreover, with doping the magnetic impurity\cite{Kikkawa}, the spin relaxation time decreases due to the 
spin flip scattering by magnetic impurities. 

However, our experimental findings indicate that the spin temperature rises slowly towards the hole and lattice temperature.
According to the three temperature model \cite{Beaurepaire}, the rising speed of spin temperature is determined by the heat capacities and the heat flow rates. The heat capacity of the spin system is 18${\rm J/K \cdot m^3}$, which is much smaller than those of lattice and hole systems, $1.0 \times 10^6{\rm J/K \cdot m^3}$ and $1.8 \times 10^2{\rm J/K \cdot m^3}$, respectively\cite{Blakemore}. 
Therefore, we need to consider the anomalous thermal isolation of spin systems. This resembles what we have observed in a half-metallic ferromagnet Sr${\rm _2}$FeMoO${\rm _6}$\cite{Kise}.

After the rapid non-radiative relaxation, the subsequent carrier scattering mainly occurs near the Fermi surface. Therefore, the demagnetization rate of carriers strongly depends on their spin polarization near the Fermi level. In particular, if the density of state for one spin direction is much smaller than that of the other direction, the spin-flipping is inefficient. 
This situation occurs for a system where the exchange splitting of carriers' band is larger than the Fermi energy, leading to the half-metallic electronic structure. Therefore, the observed slow demagnetization invokes this condition which holds for ${\rm Ga_{0.94}Mn_{0.06}As}$ as well as for Sr${\rm _2}$FeMoO${\rm _6}$.

The observed temporal responses can be numerically simulated by using the three-temperature model\cite{Beaurepaire}. The solid curve of Fig.~\ref{fig3} shows that this model can reproduce the temporal 
evolution of spin systems within a wide time interval. The rising time of the spin temperature is found to be 500ps.
It is necessary to emphasize that the obtained coupling constant between the charge and spin systems is much smaller than that in Ni.
The observed spin relaxation at longer time delay might be due to spin-lattice coupling, which stems from the magnetic anisotropy\cite{Hubner}.

In conclusion, we observe the magnetization dynamics in ${\rm Ga_{0.94}Mn_{0.06}As}$ over a wide 
temporal range from ps to ns using two-color TR-MOKE measurements. 
We discover that the spin system thermalizes much slower than the hole and lattice systems, indicating spin-charge thermal isolation. This experimental finding confirms that ${\rm Ga_{0.94}Mn_{0.06}As}$ has a half-metallic electronic structure, making it a prospective material for spintronic applications. The half metalicity indicates a large exchange energy of the carriers compared to the Fermi energy, which plays an essential role for the magnetism and transport properties of this material. 

This work is supported by a Grant-in-Aid for Scientific for COE research 
(Phase Control of Spin-Charge-Photon Coupled Systems) and a Grant-in-Aid for COE 
Research (Quantum Dot and Its Application) from the Ministry of Education, Culture, 
Sports, Science, and Technology of Japan.


\begin{thebibliography}{23}
\bibitem{aboutC} The obtained dependence of $C$ on the frequency reflects the fact that our measurements 
are performed close in the band gap region where both $f_\theta$ and 
$f_\eta$ are sensitive to the phase of reflected probe light.
\end{thebibliography}

\end{document}